# Investigation of Multi-Stage Evaporation and Wave Multiplicity of Two-Phase Rotating Detonation Waves Fueled by Ethanol

Jianghong Li[a,b], Ying Lei[b,c], Songbai Yao[b,c,*], Jingtian Yu[b,c], Jingzhe Li[a,b], Wenwu Zhang[b,c]

[a]Faculty of Mechanical Engineering and Mechanics, Ningbo University, Ningbo 315211, China

[b]Zhejiang Key Laboratory of Aero Engine Extreme Manufacturing Technology, Ningbo Institute of Materials Technology and Engineering, Chinese Academy of Sciences, Ningbo 315201, China

[c]University of Chinese Academy of Sciences, Beijing 100049, China

[*]E-mail: yaosongbai@nimte.ac.cn (S. Yao)

**ABSTRACT**

In this study, a numerical investigation based on the Eulerian-Lagrangian model is conducted to explore a rotating detonation engine (RDE) fueled by liquid ethanol. The focus is on examining the characteristic phenomena of the two-phase rotating detonation wave (RDW) caused by droplet evaporation and varying inlet conditions. To enhance the evaporation of liquid fuel, pre-heated air is used, and both liquid and pre-vaporized ethanol are simultaneously injected. The distribution of ethanol droplets reveals an initial concentration near the injection surface and accumulation in the fuel-refill zone. Here, liquid droplets gradually evaporate after absorbing latent heat from the surrounding gas. The subsequent interactions between the evaporating droplets and the RDW vary with the droplet size. For droplets with diameters of $d_0$ = 5-15 μm, after being swept by the RDW, a secondary evaporation process occurs, leading to an enlargement of the width of the reaction zone. However, the chemical reactions still predominantly take place in close proximity to the detonation front. As $d_0$ further increases, droplet evaporation persists in the post-detonation expansion zone over a long distance until the remaining droplets are fully evaporated and eventually burned in this region. The study also analyzes the extinction of rotating detonations and the emergence of new detonation waves resulting from local explosions and consequent shock collisions. It is demonstrated that variations in the diameter of injected droplets and inlet temperature can lead to different operating modes with varying numbers of RDWs.

**Keywords**: Rotating detonation; Ethanol; Two-phase detonation; Eulerian-Lagrangian model; Multi-stage evaporation; Wave multiplicity

# NOMENCLATURE

| | |
|---|---|
| $A_d$ | Surface area of droplet [m$^2$] |
| $A_s$ | Sutherland coefficient |
| $B_d$ | Spalding mass transfer number |
| $C_d$ | Drag coefficient |
| $C_{p,d}$ | Heat capacity of droplet [J/kg/k] |
| $c_{v,m}$ | Heat capacity of $m$-th species |
| $D$ | Mass diffusivity [m$^2$/s] |
| $D_d$ | Diameter of the droplet [m] |
| $\mathbf{F}_d$ | Stokesian drag force [N] |
| $h_{f,m}$ | Enthalpy of formation for species $m$ [J/kg] |
| $k$ | Thermal conductivity [W/m/K] |
| $\dot{m}_d$ | Evaporation mass transfer rate of droplet [kg/s] |
| $p_{sat}$ | Saturated pressure [Pa] |
| $\dot{Q}_c$ | Convective heat transfer rate [J/s] |
| $\mathbf{q}$ | Heat flux vector [W/m$^2$] |
| $\mathbf{s}_m$ | Mass flux vector [kg/m$^2$/s] |
| $T_d$ | Temperature of droplet [K] |
| $T_g$ | Temperature of gas [K] |
| $\mathbf{V}$ | Velocity vector of gas [m/s] |
| $\mathbf{V}_d$ | Velocity vector of droplet [m/s] |
| $\boldsymbol{\tau}$ | Viscous stress tensor [kg/m/s$^2$] |
| $\mu$ | Dynamic viscosity [kg/m/s] |
| $\dot{\omega}_m$ | Chemical reaction rate of $m$-th species [kg/m$^3$/s] |

Acronym

| | |
|---|---|
| RDE | Rotating detonation engine |
| RDW | Rotating detonation wave |
| OSW | Oblique shock wave |

## 1. INTRODUCTION

Pressure gain combustion (PGC) has gained global attention for its potential in improving the thermodynamic efficiency of propulsion systems [1-3]. PGC is achieved through detonation engines, including the pulsed detonation engine (PDE), oblique detonation engine (ODE), and rotating detonation engine (RDE). Recently, there has been a notable increase in efforts to incorporate RDEs into practical applications [4, 5].

Unlike other types of detonation engines, the RDE is characterized by the continuous propagation of one or multiple rotating detonation waves (RDWs) along the circumferential direction.

Initial investigations into RDEs primarily focused on the utilization of gaseous fuels [6-12]. However, as the research on RDEs is progressing toward practical engineering applications, there has been an increasing interest in exploring the feasibility of utilizing liquid fuels. Unlike explosive gaseous fuels, however, liquid fuels present distinct challenges due to their narrow explosive limit and long ignition delay time, rendering the initiation of detonation combustion difficult, and the necessity of combustion enhancement using oxygen-enriched air, pre-heated hot air or hydrogen addition is needed [13-20]. With the ever-growing global concern regarding climate change and the urgent need to reduce greenhouse gas emissions, recently there is also a trend to use cleaner and more sustainable fuel alternatives for RDEs. For example, Huang et al. [21] investigated the potential use of ammonia for the RDE. The experiment was conducted in a hollow chamber with a Laval nozzle and was indicated that ammonia-fueled rotating detonations could be achieved with an oxygen mass fraction of at least 43%, but the operating range decreased as the oxygen mass fraction decreases from 83% to 43%. In the numerical study of Sun et al. [22], the flow field and propagation modes of the RDE fueled by hydrogen-enriched ammonia and oxygen-enriched air were investigated. The simulations also showed that stable detonations occurred only when the hydrogen concentration is at least 0.2 (mole fraction), with increased hydrogen concentration leading to a higher wave velocity and decreased height of the detonation wave. Furthermore, the study suggested that maximizing cycle efficiency required an optimal hydrogen concentration of 0.3 and an equivalence ratio of 0.8. In the study of Wang et al. [23], ammonia/hydrogen/air rotating detonations were achieved at high inlet total temperatures and the pressure gain performance was evaluated at different fuel ratios.

Ethanol is another renewable fuel option for the RDE, offering advantages such as lower toxicity and easier handling compared to ammonia. In a recent study by Yoneyama et al. [24], the successful implementation of rotating detonation combustion in a cylindrical combustor using liquid ethanol and gaseous oxygen was experimentally demonstrated. The research confirmed the presence of both detonation combustion and deflagration combustion within the studied operational range of mass flow rates (26-40 g/s) and equivalence ratios (0.4-1.7). In addition, it was shown that an increase in the equivalent ratio resulted in higher values for both thrust and specific impulse [25]. In our recent study, we unveiled the structure of a heterogeneous two-phase RDW fueled by hydrogen-enriched ethanol and air. Despite the favorable attributes, the potential use of ethanol in RDEs presents certain challenges. Ethanol's lower energy density compared to conventional fossil fuels necessitates further investigations into optimizing fuel injection, combustion enhancement, and maintaining self-sustained two-phase rotating detonations. But there remains a scarcity of studies investigating the comprehensive properties of liquid-fueled RDEs.

Building upon our previous investigation [26, 27], we present a more comprehensive analysis of the fundamental characteristics of the two-phase RDW, where ethanol is used as an alternative to conventional fossil fuels such as kerosene. Specifically, the investigation involves numerical simulations utilizing the Eulerian-Lagrangian model to elucidate the different stages of interactions between evaporating droplets and the RDW, while analyzing the phenomena of pre-ignition, local explosions, extinctions, and RDW formation under diverse operating conditions. The structure of this paper is organized as follows: the initial section presents the numerical methods employed in our simulations, including the presentation of the governing equations, the density-based solver, chemical kinetics, boundary conditions, and mesh discretization. The subsequent section constitutes the main body of the paper, where we thoroughly discuss and summarize the results obtained from our simulations. Lastly, the concluding section provides a comprehensive presentation of the key findings derived from this study.

## 2. Methods

### 2.1 Governing equations

#### 2.1.1 Gas phase

The compressible Navier-Stokes equations are employed to solve for the carrier phase, along with the transport equations for the reacting species, assuming unity Lewis and Schmidt numbers [23, 28],

$$\frac{\partial \rho}{\partial t} + \nabla \cdot [\rho \boldsymbol{V}] = S_m, \quad (1)$$

$$\frac{\partial (\rho \boldsymbol{V})}{\partial t} + \nabla \cdot [(\rho \boldsymbol{V}\boldsymbol{V})] = -\nabla p - \nabla \cdot \boldsymbol{\tau} + S_u, \quad (2)$$

$$\frac{\partial (\rho E)}{\partial t} + \nabla \cdot [(\rho E + p)\boldsymbol{V}] = \nabla \cdot [\boldsymbol{\tau} \cdot \boldsymbol{V}_g] - \nabla \cdot \boldsymbol{q} + S_e, \quad (3)$$

$$\frac{\partial(\rho Y_m)}{\partial t} + \nabla \cdot [(\rho Y_m \boldsymbol{V})] = -\nabla \cdot \boldsymbol{s_m} + \dot{\omega}_m + S_Y, \tag{4}$$

where $\boldsymbol{\tau} = \mu\left((\nabla \boldsymbol{U}) - (\nabla \boldsymbol{U})^T - \frac{2}{3}(\nabla \cdot \boldsymbol{U})\boldsymbol{I}\right)$ is the viscous stress tensor. The dynamic viscosity $\mu = A_s \sqrt{T}/(1 + \frac{T_s}{T})$ is computed using Sutherland's law [29], where the coefficients are given by $A_s = 1.67212 \times 10^{-6}$ kg/m·s·$\sqrt{K}$ and $T_s = 170.672$ K. The diffusive heat flux is given by Fourier's law $\boldsymbol{q} = -k\nabla T$ where the thermal conductivity $k$ is evaluated using the Eucken correlation [30], and $\boldsymbol{s}_m = -D\nabla(\rho Y_m)$ is the mass flux vector with the mass diffusivity computed as $D = k/\rho C_p$. As discussed in our previous study [26, 27] and in line with refs. [23, 31, 32], turbulence models are not implemented here and the Navier-Stokes equations are directly solved. Moreover, $\dot{\omega}_m$ is the net production rate of the $m$-th species due to chemical reactions. $S_m$, $S_u$, $S_e$, and $S_Y$ are the source terms related to the exchanges of mass, momentum, energy, and species between the gas and liquid phases and will be elaborated upon in the subsequent discussion.

**2.1.2 Liquid phase and evaporation model**

The dynamics of the Lagrangian droplets under the point-source assumption are described by the following equations,

$$\frac{dm_d}{dt} = -\dot{m}_d, \tag{5}$$

$$\frac{d\boldsymbol{V}_d}{dt} = \frac{\boldsymbol{F}_d}{m_d}, \tag{6}$$

$$C_{p,d}\frac{dT_d}{dt} = \frac{\dot{Q}_d - \dot{m}_d h(T_d)}{m_d}, \tag{7}$$

where $C_{p,d}$ is the heat capacity of the droplet, and $h(T_d)$ is the fuel vapor enthalpy on the droplet surface. $m_d$ is the mass of the droplet and the evaporation mass transfer rate $\dot{m}_d$ is computed according to Abramzon and Sirignano [33]

$$\dot{m}_d = \pi D_d Sh D_{vap}\rho_f \ln(1 + B_m), \tag{8}$$

where $B_m = \frac{X_{vap,s} - X_{vap,g}}{1 - X_{vap,g}}$ is the Spalding mass transfer number, $D_{vap}$ is the vapor mass diffusivity, and $\rho_s$ is the density of the film around the droplet. $X_{vap,s}$ and $X_{vap,g}$ are the mole fractions of fuel vapor on the droplet surface and in the ambient gas. According to Raoult's law,

$$X_{vap,s} = X_{vap,g}\frac{p_{sat}}{p} \tag{9}$$

where $p_{sat}$ is the saturated vapor pressure. Noted that the evaporation of droplets can also be described by non-equilibrium models [34, 35], and the deviation between equilibrium and non-equilibrium models was discussed in ref. [36]. Miller et al. [37] conducted a comprehensive analysis of the equilibrium and non-equilibrium models and stated that the non-equilibrium models could be more accurate at high evaporation rates; however, it was also highlighted that the contribution of the non-equilibrium effects to the overall behavior of droplet evaporation were relatively nonessential when compared to the experimental data. This is still an open question and necessitates further investigation. In accordance with the previous studies on two-phase RDE [28, 38-40], here a liquid-vapor equilibrium is assumed and the vapor partial pressure on the droplet surface is determined by the Clausius-Clapeyron law. The Sherwood number $Sh$ is calculated by

$$Sh = 2.0 + 0.6 Re_d^{\frac{1}{2}} S_c^{\frac{1}{3}} \tag{10}$$

using the empirical Ranz-Marshall correlation [41, 42].

The Stokesian drag force $\mathbf{F}_d$ in Eq. (6) for spherical droplets is computed by

$$\mathbf{F}_d = \frac{18\mu}{\rho_d D_d^2} \frac{C_d Re_d}{24} m_d (\mathbf{V} - \mathbf{V}_d) \tag{11}$$

where $C_d$ is the drag coefficient [43]:

$$C_d = \begin{cases} 0.424 & , Re_d > 1000 \\ \frac{24}{Re_d}\left(1 + \frac{1}{6} Re_d^{\frac{2}{3}}\right) & , Re_d < 1000 \end{cases} \tag{12}$$

In Eq. (7), $\dot{Q}_d = h_c A_d (T - T_d)$ is the convective heat transfer rate, where the convective heat transfer coefficient $h_c$ is determined using the correlation by refs. [41, 42], and $h_{vap}$ is the enthalpy of the fuel vapor.

**2.1.3 Eulerian-Lagrangian coupling**

The carrier phase and the Lagrangian droplets are then two-way coupled through the source terms in Eqs. (1-4), i.e.,

$$S_m = \frac{1}{V} \sum_{N_d} \dot{m}_d, \tag{13}$$

$$S_u = -\frac{1}{V} \sum_{N_d} \mathbf{F}_d, \tag{14}$$

$$S_e = -\frac{1}{V} \sum_{N_d} (\dot{Q}_d - \dot{m}_d h_{vap}), \tag{15}$$

$$S_Y = \begin{cases} S_m, & \text{fuel species} \\ 0, & \text{other species} \end{cases} \quad (16)$$

where $V$ is the volume of the single cell and $N_d$ is the number of droplets in the cell.

### 2.1.4 Reaction mechanism

The chemical reaction of ethanol-air is modeled using Westbrook and Dryer's global single-step mechanism [44],

$$C_2H_5OH + 3O_2 \xrightarrow{k_1} 2CO_2 + 3H_2O \quad (17)$$

The reaction rate is computed according to the Arrhenius law

$$k_1 = AT^n \exp\left(-\frac{E_a}{RT}\right)[\text{Fuel}]^a[\text{Oxidizer}]^b$$

where the reaction rate parameters are summarized in Table 1.

Table 1 Reaction rate parameters for ethanol oxidation reaction (units in cm-sec-mole-kcal-K) [44].

| Fuel | $A$ | $n$ | $E_a$ | $a$ | $b$ |
|---|---|---|---|---|---|
| $C_2H_5OH$ | $1.5 \times 10^{12}$ | 0.0 | 30.0 | 0.15 | 1.6 |

### 2.2 Numerical setup

The simulations in this study utilize the finite volume method, which is implemented in the open-source OpenFOAM library (v2206) [45]. In order to accurately capture shock phenomena, a compressible solver is developed based on the central-upwind Kurganov and Tadmor (KT) schemes [46], and a Lagrangian particle tracking library is integrated. The effectiveness of the KT schemes in resolving detonation waves has been extensively validated in numerical studies of RDEs [28, 31, 32, 47].

The liquid droplets are assumed to be in the dilute regime, where considerations regarding inter-droplet interactions and droplet break-up are neglected. The droplets are introduced in parcels at the inlet, with the same velocity as the gaseous phase. A non-reflecting boundary is applied at the outlet, where the far-field conditions are set to the ambient temperature ($T_\text{inf} = 300 \text{ K}$) and pressure ($p_\text{inf} = 1.013 \times 10^5 \text{ Pa}$). The ethanol droplets are assumed to be pre-atomized, matching the velocity of the local gaseous flow. For the injection of the gaseous flow, a commonly used approach in RDE simulations is employed [23, 27, 28, 31, 47]. This approach generates an injection flow based on both the inlet total pressure of the reservoir and the local pressure of the computation cell on the headwall. The computational domain of the rotating detonation combustor (RDC) is simplified into a two-dimensional representation, under the assumption that the radial dimension of the RDC is considerably smaller (by roughly one order of magnitude) in comparison to its circumferential and axial dimensions. The RDC is unwrapped along the circumferential direction, forming a computational domain ($2\pi R \times L$) with dimensions of 160 mm × 100 mm (refer to Fig. 1). Hence, the effects of wall curvature [12, 48-50] and the influence of near-wall heat flux and turbulence [51, 52] are neglected.

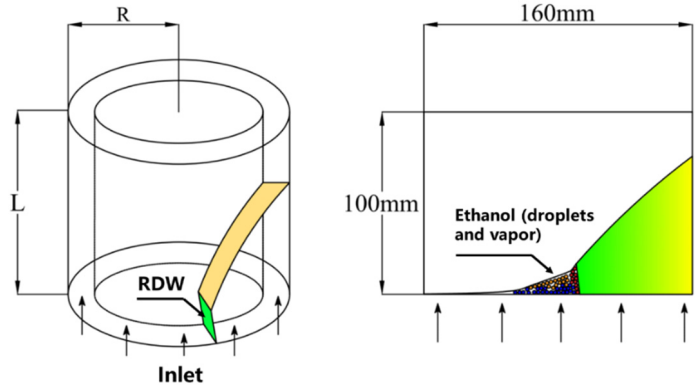

Figure 1 Configuration of the RDE and the computational domain.

## 3. Results and discussion

### 3.1 Simulation cases

Previous studies on liquid-fueled detonations have highlighted the difficulties associated with initiating detonations using liquid fuels, especially for environmentally friendly fuels such as ethanol, which have a lower energy density compared to fossil fuels. Experimental investigations on liquid-fueled detonations in tubes have reported that the presence of vapor played a crucial role in initiating and sustaining the self-sustained propagation of liquid-fueled detonations [53]. Therefore, in this study, the liquid ethanol is injected together with pre-vaporized ethanol vapor (60% by mass) to support the propagation of the ethanol-fueled RDW, with a global equivalence ratio of approximately 1.0.

Additionally, we use air instead of oxygen-enriched air as the oxidizer, which is supplied at an inlet total temperature of 900 K to fascinate the evaporation of liquid ethanol. To investigate the effects of inlet total temperature, additional simulation cases with $T_0$ increased to 1000 K and 1200 K are conducted (Cases C-D), and the effects of inlet total temperature will be discussed in detail later. The inlet total pressure $p_0$ is fixed at 20 atm in all cases. The diameter of the droplet varies from 5 to 25 μm. Case A is a simulation case fueled by a premixed stoichiometric ethanol-air mixture at $T_0=$ 900 K, which will result in a gaseous RDW similar to that in the study of ref. [54], and will be used for comparison with the two-phase RDW. Table 2 summarizes the operating conditions of the simulation cases.

Table 2 Inlet and initial conditions of the simulation cases.

| Case | Mass flow rate of ethanol (kg/s) | $\phi$ | $T_0$ (K) | $d_0$ (μm) |
|---|---|---|---|---|
| A | 0.047 (gaseous ethanol-air mixture) | 1.0 | 900 | - |
| B1 | 0.025 | ≈1.0 | 900 | 5 |
| B2 | | | | 10 |
| B3 | | | | 15 |
| B4 | | | | 20 |
| B5 | | | | 25 |
| B6 | 0.04 | ≈1.4 | 900 | 5 |
| C1 | 0.025 | ≈1.0 | 1000 | 5 |
| C2 | | | | 10 |
| C3 | | | | 15 |
| D1 | | | 1200 | 5 |
| D2 | | | | 10 |
| D3 | | | | 15 |
| D4 | | | | 20 |

## 3.2 Validations

### 3.2.1 Chemical kinetics for detonation simulations

To assess the applicability of the one-step chemistry model for detonation simulations, numerical experiments are conducted in detonation tubes using ethanol (vapor) – air mixtures. These tubes have dimensions of 500 mm in length and 80 mm in width and are initially filled with premixed ethanol-air mixtures spanning different equivalence ratios. The cell size of the ethanol-air mixture is then computed and compared against experimental data obtained from Diakow et al. [55]. In their experimental investigation, the measured ethanol-air detonation cell width ranged between 30 and 40 mm and exhibited small variation with the mixture's composition, as depicted in Fig. 2. Under the same initial conditions of $p_0 = 0.1$ MPa and $T_0 = 373$ K and 310 K, the average cell sizes obtained from the simulations also demonstrate consistent behavior as the mole fractions of ethanol change within the mixture. Though the simulation results cannot precisely replicate the exact cell size of the ethanol-air detonation observed in the experiments, the computed cell sizes in general align within an acceptable range and demonstrate magnitudes of cell size that are comparable.

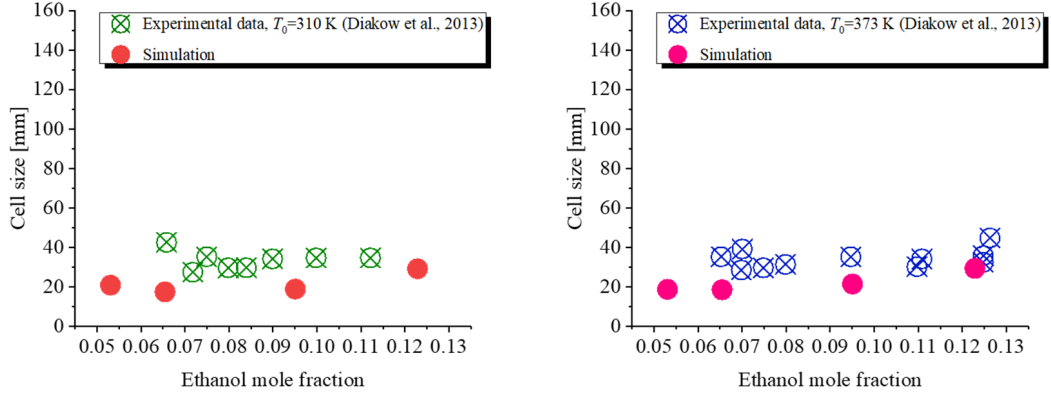

Figure 2 Detonation tube simulations with premixed ethanol-air at $p_0 = 0.1$ MPa and $T_0 = 310$ K and 373 K. The cell size measurements are plotted according to the experimental study of Diakow et al. [55].

### 3.2.2. Grid sensitivity analysis

The circumferential direction in Fig.1 is uniformly divided into mesh cells with a grid spacing of $\Delta x = 100$ μm. However, in order to capture the details near the front of the RDW, the mesh cells are stretched along the axial direction with a stretching factor of 10 (bottom to top). This ensures a finer mesh resolution in the region where a more refined mesh is necessary. Figure 3 displays snapshots of temperature contours of the benchmark (Case A) at three different mesh sizes, i.e., the total number of grid nodes ranges from $2 \times 10^5$ to $3.2 \times 10^6$. The results confirm that the flow fields and RDW structures remain largely unchanged across the resolutions. Although a finer mesh provides more detailed information about the RDW front and the post-RDW region, no fundamental differences are observed. Additionally, the heights of the RDWs, which are crucial detonation parameters, are approximately the same for all cases. To assess the accumulation of computational errors, the proposed method by ref. [56] for chemical reacting flows is used. The total accumulation errors $\bar{S}_{\text{err}} = S_{\text{err}} \cdot \sqrt{n}$ are computed across simulations conducted at various resolutions. Here, $S_{\text{err}}$ represents the summation of relative errors of integration in two dimensions and $n$ denotes the total number of time steps. For the simulation cases to reach a quasi-steady state, the total accumulation errors $\bar{S}_{\text{err}}$ are approximately 0.013% and 0.0002% for the coarsest and finest meshes, respectively, which are well within the allowable limits. Furthermore, to satisfy the point-source assumption, it is necessary for the mesh size to be significantly larger than the size of the Lagrangian droplets. Taking these factors into account, the total number of grid nodes is determined to be $8 \times 10^5$ for the subsequent simulation cases, ensuring the fulfillment of the coupling assumption between the gas phase and Lagrangian droplets.

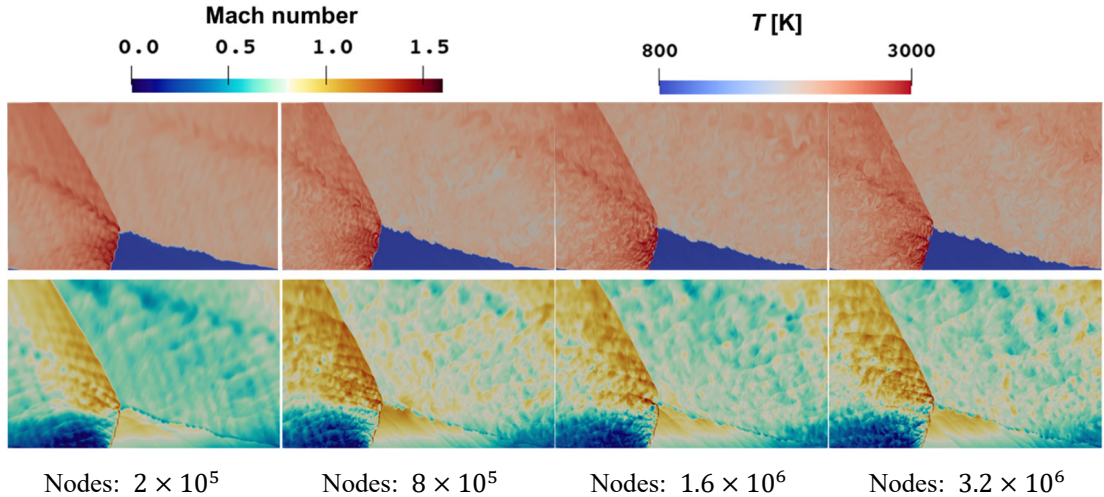

Figure 3 Simulations of the ethanol (vapor)-air RDWs under the three different resolutions. Premixed ethanol-air vapor at $T_0 = 900$ K, $p_0 = 0.1$ MPa (Case A).

### 3.3. Primary structure of the ethanol-air RDW

Figure 4 presents a comparative analysis between Case B1 at $d_0$=5 μm and Case A. Notably, with smaller droplet sizes, the two-phase ethanol-air RDW exhibits similarities to the premixed kerosene vapor-air fueled RDW. This resemblance is evident in the overall structure of the detonation front, oblique shock wave (OSW), and shear layer. These findings support the conclusions drawn from our previous investigation on kerosene-fueled RDWs [27]. However, distinct differences exist due to droplet evaporation and subsequent interactions with the RDW.

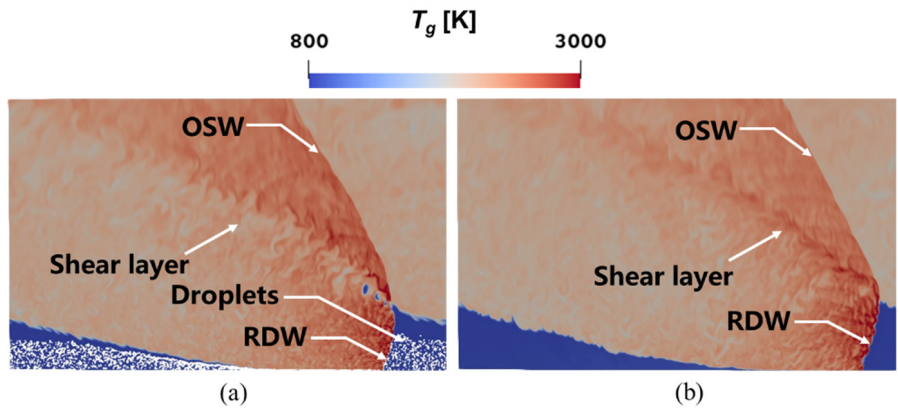

Figure 4 Overall structure of (a) two-phase ethanol-air RDW and (b) gaseous RDW with a premixed ethanol (vapor)-air mixture at $T_0 = 900$ K.

For the two-phase RDW, the initial stage of droplet evaporation starts in the fuel-refill zone. In Fig. 5, the distribution of ethanol in the flow field is presented on the right side, while the left side provides the plots of the temperature and ethanol vapor mass fraction ($Y_{C_2H_5OH}$) variations along a specific line along the axial direction. Initially, a certain amount of ethanol vapor is present near the injection surface due to the simultaneous injection of ethanol vapor with liquid droplets. Then, as the liquid ethanol droplets undergo evaporation upon injection, ethanol vapor accumulates in the fuel-refill zone. The evaporation mass transfer rate of the droplet $\dot{m}_d$ computed by Eq. (5) is presented in Fig. 6, and it is shown that the evaporation of liquid droplets occurs at a minimal rate upon initial injection. However, as the droplets progress downstream into the fuel-refill zone, a notable increase in $\dot{m}_d$ becomes evident. The mass fraction of ethanol vapor then attains its maximum value downstream at the height of the RDW. As the RDW progresses through the region, the residual droplets experience swift evaporation and are depleted. Due to the latent heat of evaporation absorbed by the liquid droplets, the temperature along the axial direction of the refill zone in Fig. 5a exhibits a slight decline before reaching its peak at the interface between the burned products and the fuel-refill zone, compared to the uniform temperature distribution of the refill zone in Fig. 5b.

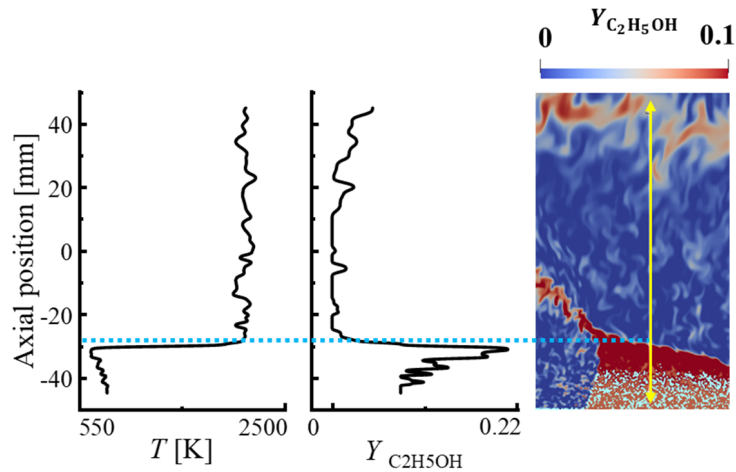

(a) Two-phase RDW

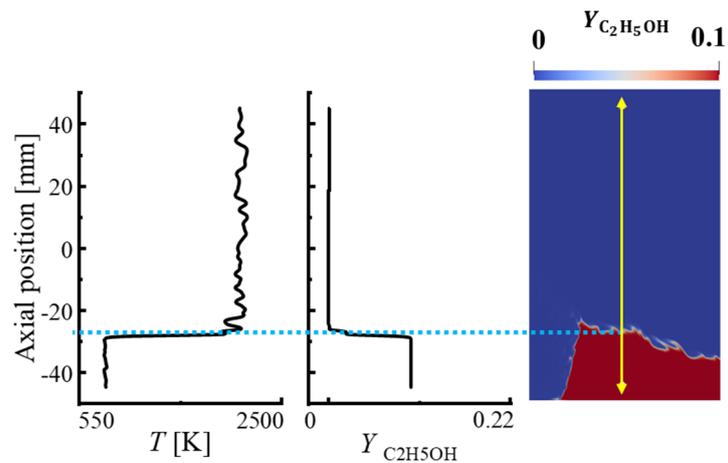

(b) Gaseous premixed RDW

Figure 5 Axial variations of temperature and ethanol vapor mass fraction.

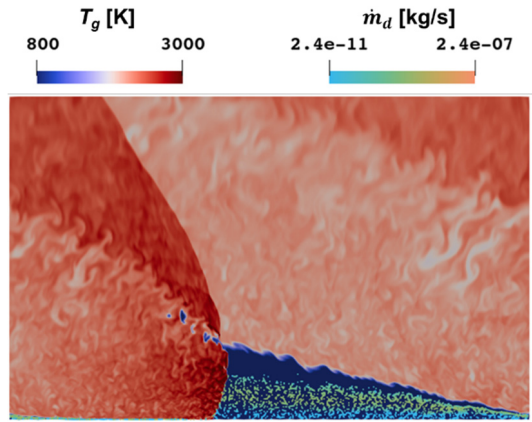

Figure 6 Contours of the evaporation mass transfer rates of droplets and temperature of the surrounding gas.

To illustrate the subsequent interactions with the detonation front, which are found to be a two-stage process, the variations of the heat release rates of chemical reactions (per unit volume) $\dot{Q}$ along the circumferential direction (denoted by the arrow line) are presented in Fig. 7. The values of $\dot{Q}$ of the two-phase and gaseous RDWs reach their peak values near the fronts at $6.93 \times 10^{13}$ J·m$^{-3}$·s$^{-1}$ and $5.14 \times 10^{13}$ J·m$^{-3}$·s$^{-1}$, respectively, indicating a higher strength of the gaseous RDW. Figure 7 indicates that the reaction zone of the detonation front of the two-phase RDW is still as sharp as that of the gaseous RDW. However, as the droplet size increases, after the reactants are swept by the detonation front of the two-phase RDW, the completion of reactions requires more time, and the reaction zone of the detonation front has an enlarged width.

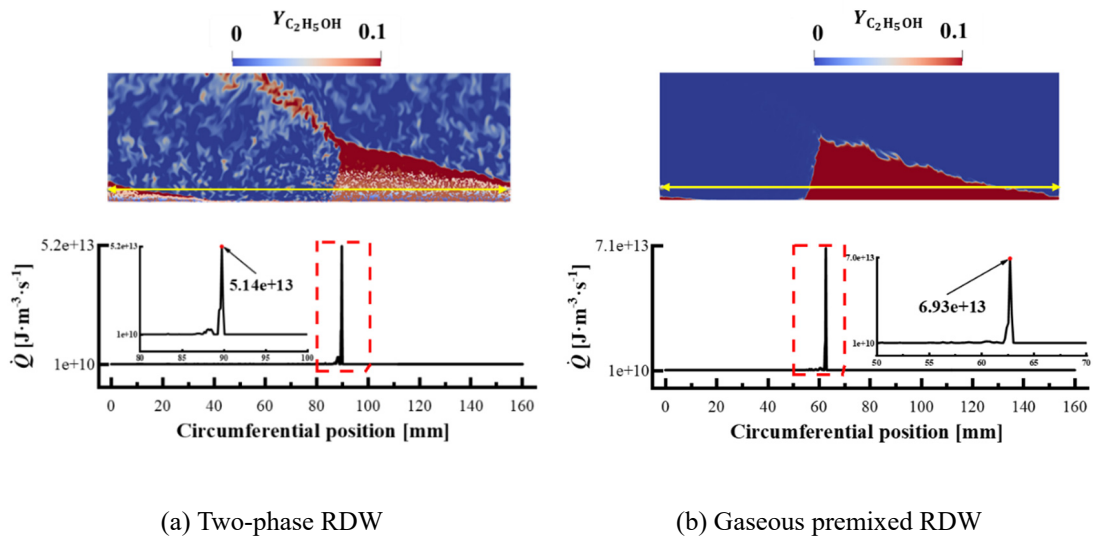

(a) Two-phase RDW  (b) Gaseous premixed RDW

Figure 7 Circumferential variation of temperature and ethanol vapor mass fraction ($Y_{C_2H_5OH}$).

This phenomenon is evident in Cases B2 and B3 where the diameter of the injected ethanol droplets increases to 10 and 15 μm. The unconsumed droplets undergo a secondary-stage evaporation before being fully consumed by a secondary reacting front, resulting in the formation of a dual-front structure previously reported in our previous studies [26, 27]. The pre-heating effect from the hot air enhances the vaporization of upstream droplets upon being swept by the detonation wave. In contrast, the just-injected ethanol droplets require a longer time for complete evaporation, leading them to travel further beyond the RDW. Consequently, a skew triangular region is formed by the unburned ethanol droplets in conjunction with the detonation wave front (see the left side of Fig. 8a). Figures 8b and 8c represent the contours of $\dot{m}_d$ and a close-up of $\dot{Q}$ in the near-front region. It can be seen that the heated droplets passing through the detonation front will continue to evaporate inside this skew triangular region. At $d_0 \leq 15$ μm, the completion of droplet evaporation and reactions still mainly occur near the detonation front after the above-mentioned two-stage evaporation.

In contrast, as the diameter of the injected ethanol droplets increases to 20 μm and 25 μm, the triangular region loses its structural stability, and a substantial number of droplets within the region fail to evaporate completely. These droplets continue to travel even further into the expansion zone, covering a considerable distance, until they are ultimately evaporated by the high-temperature products. As a result, the aforementioned skewed triangular region collapses as droplets travel progressively farther from the reacting front.

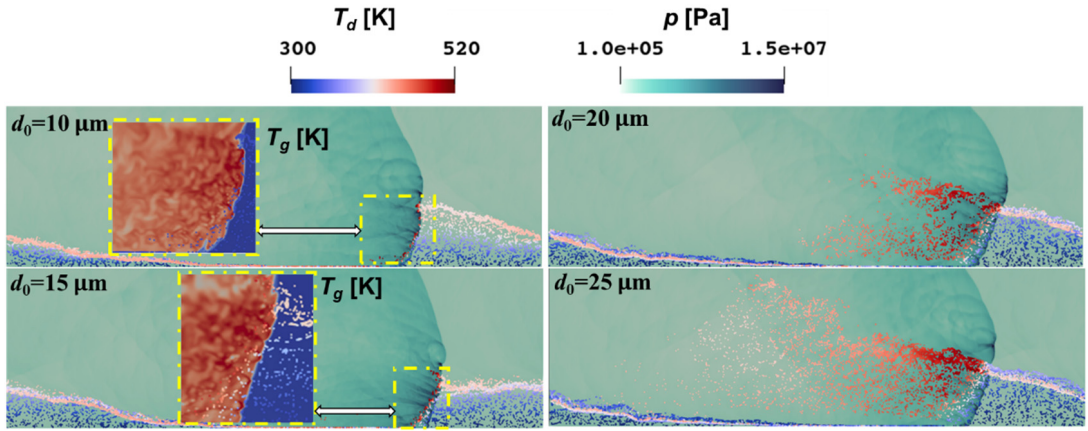

(a)

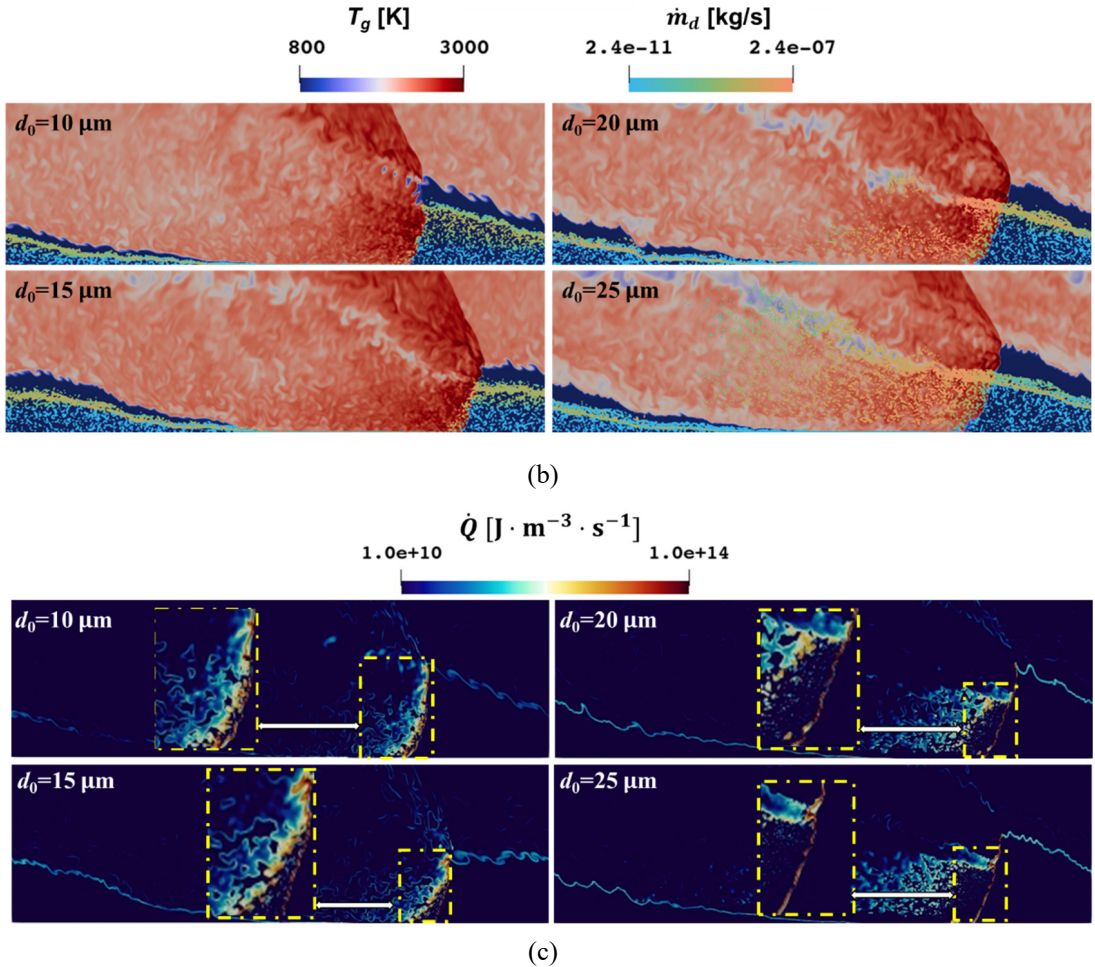

(b)

(c)

Figure 8 Evaporation and reactions of ethanol droplets at different initial sizes in the two-phase RDW flow field.

Figure 9 presents the heat reaction rates $\dot{Q}$ along the circumferential direction of the RDWs, similar to the plot shown in Fig. 7. The results reveal that at the droplet diameter of $d_0$=10 μm, the peak value of $\dot{Q}$ is followed by multiple secondary reactions that occur almost simultaneously with the secondary evaporation process. On the other hand, for $d_0$=15 μm, there is a noticeable gap between the position of the peak value of $\dot{Q}$ and the appearance of the secondary peak. This gap corresponds to the secondary-stage evaporation zone where droplet evaporation continues, and the occurrence of secondary reactions is not as significant as that in the case with $d_0$=10 μm. As $d_0$ increases to 20-25 μm, the droplets travel to a greater distance from the detonation front into the expansion region, the secondary-stage evaporation zone becomes considerably wider, and secondary reactions mainly occur in the post-detonation expansion zone instead of near the detonation front.

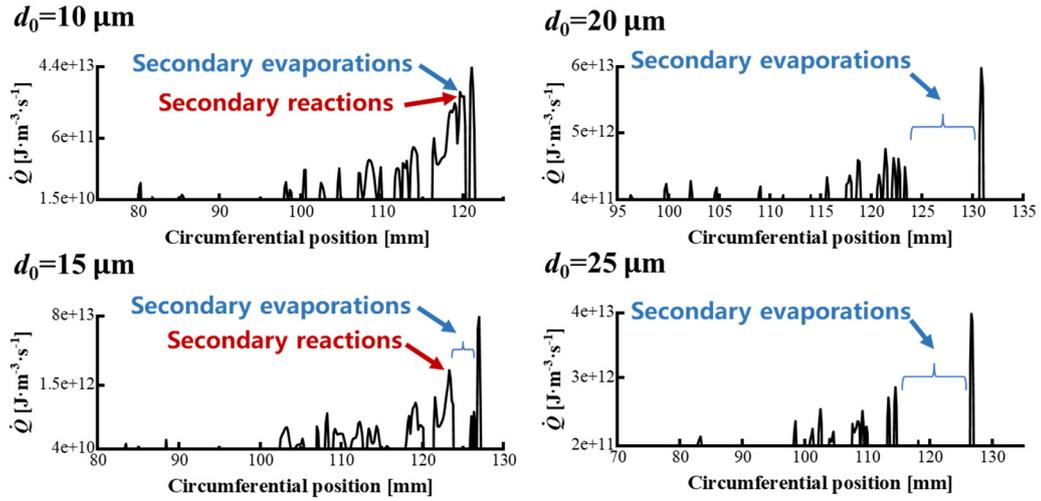

Figure 9 Circumferential variation of heat release rates in the two-phase RDW flow field with droplets of varying sizes.

Figure 10 illustrates the temporal evolution of the flow field of the two-phase RDW after we increase the mass flow rate of ethanol of Case B1 from 0.025 kg/s to 0.04 kg/s (Case B6), corresponding to a fuel-rich condition. The RDW remains basically stable at $t=370$ μs. However, as the RDW continues propagating, the accumulation of excessive ethanol droplets in the fuel-refill region gives rise to shock waves above the RDW. These shock waves undergo further development and trigger localized explosions due to the presence of high-temperature ethanol vapor in the adjacent region, as indicated in Fig. 10. The plots demonstrate the collision that generates both forward- and backward-propagating shock waves. The forward-propagating shock wave continues to evolve and transforms into a new RDW by utilizing the fuel mixture in the refill region. In contrast, the backward-propagating shock wave gradually weakens during propagation and does not initiate the formation of a new RDW (see Fig. 11).

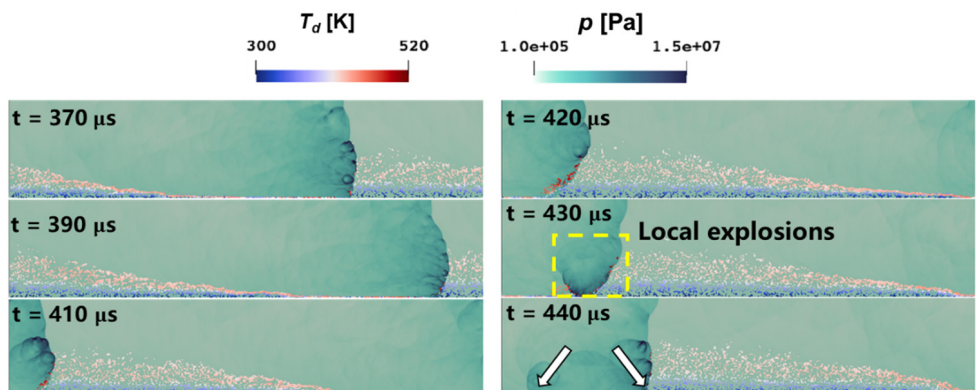

Figure 10 Pressure and droplet temperature contours illustrating local explosions resulting from increased mass flow rate of ethanol.

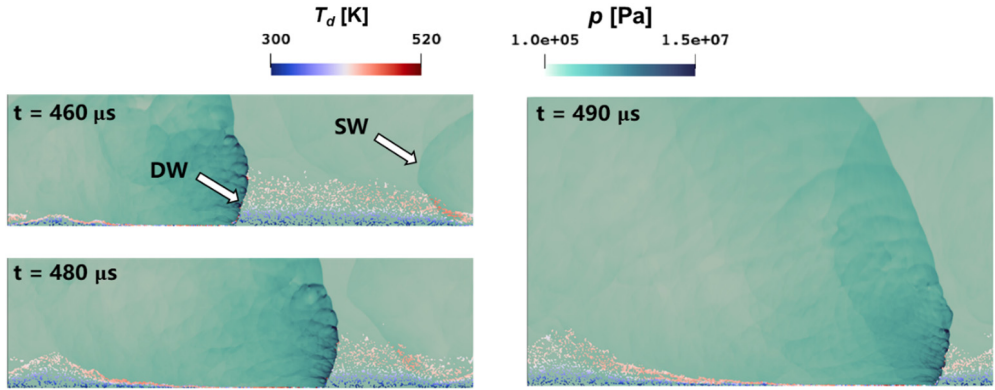

Figure 11 Pressure and droplet temperature contours illustrating the collision between the leading RDW and the generated backward-propagating shock wave.

The aforementioned collisions and re-stabilization events persist throughout the propagation of the RDW, resulting in periodic weakening and strengthening of the RDW. Eventually, at $t$ =580 μs (Fig. 12), the RDW loses its ability to completely consume the liquid droplets ahead of it. Moreover, the decrease in temperature caused by the release of latent heat from the vaporizing droplets leads to the decoupling and failure of the RDW, as shown in Fig. 12.

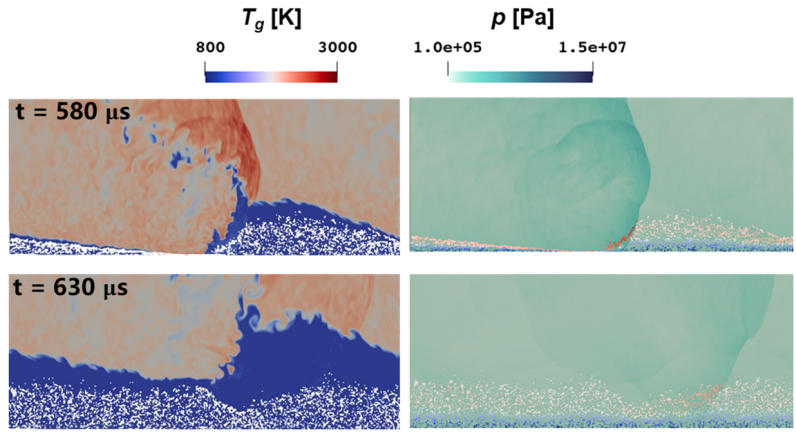

Figure 12 Temperature (gas phase) and pressure contours illustrating the extinction of the RDW as a consequence of collisions.

### 3.4 Effects of inlet total temperature

When the inlet total temperature $T_0$ is raised to 1000 K, i.e. Case C series, the stabilized flow field of the RDWs demonstrates similarities to that of the Case B series. As shown in Fig. 13, in the flow field of Case C1, the majority of ethanol droplets undergo temperature elevation and evaporation as they are swept by the detonation wave. In Case C2, there are a few high-temperature unburned ethanol droplets present in the post-detonation region, and this phenomenon becomes more pronounced as the diameter of the ethanol droplets increases. In Case C3, a significant number of high-temperature unburned ethanol droplets are observed in the post-detonation region, indicating that with the increase in droplet diameter, the transition of ethanol droplets from liquid to gas phase becomes more challenging. However, compared to the Case B series, the increase in inlet total temperature accelerates the evaporation rate of the droplets. As a result, when the droplets are relatively large, they penetrate through the detonation wave and concentrate primarily near the downstream intake wall in the expansion zone.

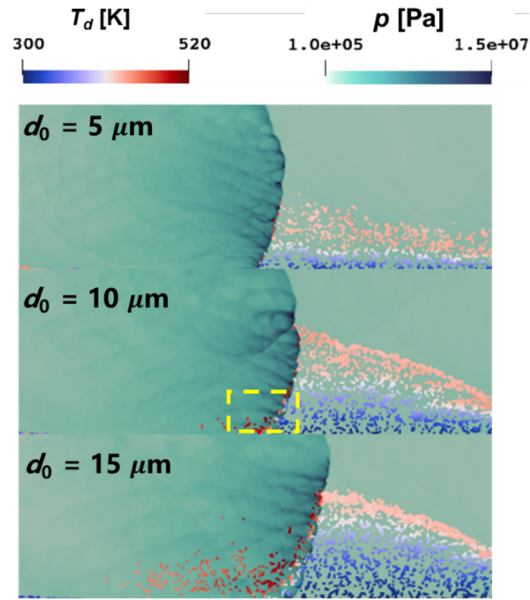

Figure 13 Stabilized flow fields of the two-phase RDW at $T_0$ = 1000 K, $d_0$=5-15 μm. Contours of pressure and droplet temperature.

It can be seen from previous analysis that the liquid droplets will be heated by the hot air before being swept by the RDW. Therefore, the inlet total temperature has significant effects on the evaporation of the ethanol droplets. In Fig. 14, the inlet total temperature of Case B1 is increased to 1200 K, i.e. Case D1. Noted that in the fuel-refill zone, there are remaining droplets previously injected at $T_0$= 900 K. As the liquid ethanol droplets continue to be injected from the bottom surface, the newly injected droplets heated by the hot air of $T_0$=1200 K evaporate more quickly than those previously injected at $T_0$=900 K. As the high-temperature droplets move further downstream, they start to evaporate into ethanol vapor before being swept by the RDW, as shown in Fig. 14.

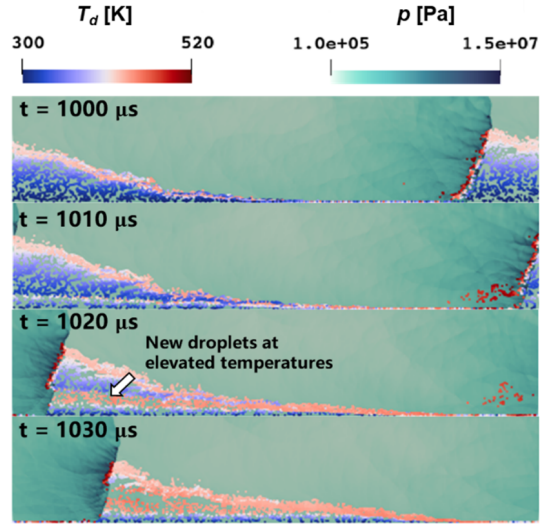

Figure 14 Injection and evolution of droplets at the elevated temperature of $T_0 = 1200$ K. Contours of pressure and droplet temperature.

In the fuel-refill zone, a small fraction of ethanol vapor undergoes ignition before being swept by the RDW, leading to localized high temperatures (Fig. 15). A substantial quantity of ethanol vapor is combusted consequently. As the RDW progresses and encounters the mixture of ethanol vapor and droplets ahead, localized explosions occur, promptly consuming the adjacent fuel mixture and subsequently causing the RDW to quench.

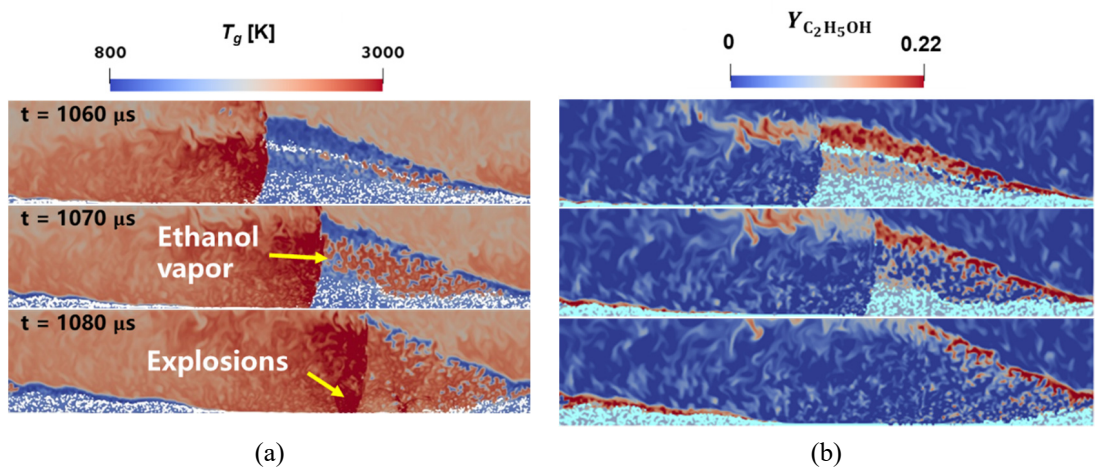

(a)          (b)

Figure 15 Contours of temperature (gas phase) and ethanol vapor mass fraction illustrating self-ignitions occurring in the fuel-refill zone.

On the other hand, due to the occurrence of local explosions, both forward- and backward-propagating shock waves are generated, causing subsequent collisions between them. With the continuous supply of fuel, some of these shock waves have the potential to develop into new detonation waves with the support of the fuel mixture, as indicated by the marked detonation waves in Fig. 16. However, the flow field fails to achieve re-stabilization, and the generation, extinction, and collisions of detonation waves continue persistently. This is because the elevated inlet temperature and the small droplet size contribute to a fast evaporation rate and rapid accumulation of ethanol vapor in the fuel-refill zone, facilitating pre-ignition and leading to subsequent intense local explosions.

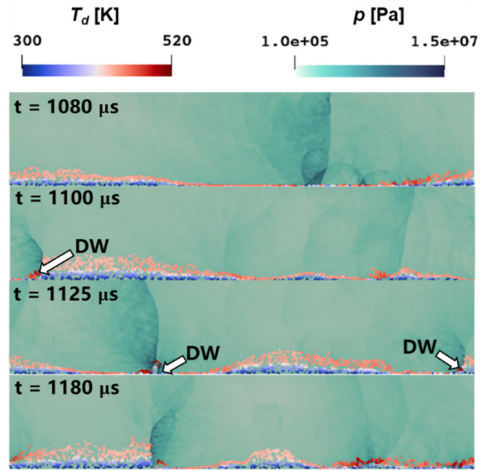

Figure 16 Pressure and droplet temperature contours illustrating the generation of new detonation waves (DWs).

That said, in Case D2, the flow field exhibits a distinct behavior as the diameter of the droplets increases to 10 μm. The flow field undergoes a self-adjusting after local explosions occur, and transitions into a double-wave mode. The subsequent analysis delves into the detailed process.

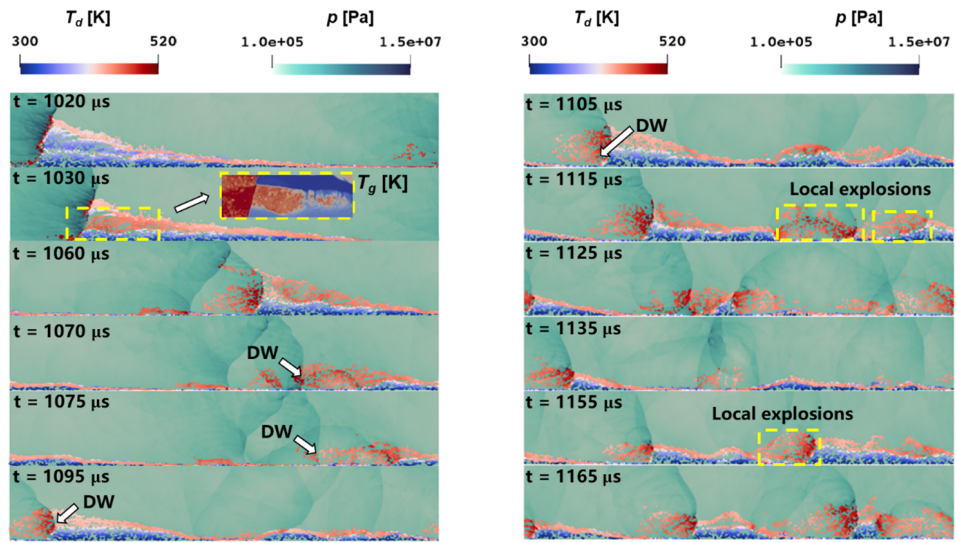

Figure 17 Pressure and droplet temperature contours illustrating the initial stage of the flow field after $T_0$ rises to 1200 K (Case D2).

The initial stage is to Case D1, i.e., as the total inlet temperature is increased from 900 K to 1200 K, the ethanol droplets ahead of the RDW rapidly evaporate, resulting in the formation of ethanol vapor that will subsequently mix with air and undergoes localized autoignition due to the high temperature. After being swept by the detonation wave, this leads to local explosions and the formation of shock waves, as illustrated in Fig. 17. These localized shock waves interact with newly injected ethanol droplets in the combustion chamber, causing further local explosions. Simultaneously, some of the shock waves evolve into new detonation waves. Due to the presence of propagating shock waves in both forward and backward directions, continuous collisions occur between the newly formed detonation waves and the shock waves in the flow field, bringing about intermittent changes in the strength of the detonation waves, and the extinction of the detonation wave also occurs, as shown in Fig. 18. Eventually, the flow field retains only two detonation waves.

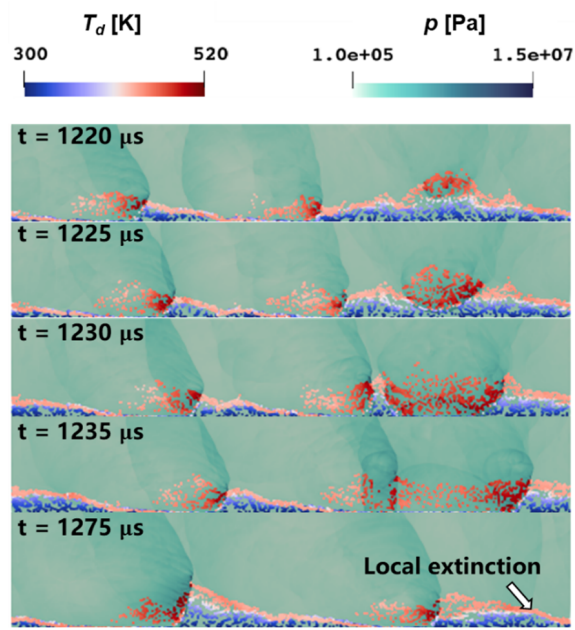

Figure 18 Pressure and droplet temperature contours illustrating the transition of the flow field into a double-wave mode (Case D2).

On the other hand, Case D3 and D4 stabilize at a three-wave operating mode. The initial phase is similar to Case D2, that is, the ignitions of ethanol vapor in the fuel-refill zone lead to the formation of new detonation waves. However, as a result of the chaotic development involving detonation wave collisions, extinctions, and generations, three self-sustained RDWs are eventually achieved, as shown in Fig. 19. This observation highlights the inherent chaotic characteristics of RDWs, a phenomenon that has been previously reported in other studies [31, 57].

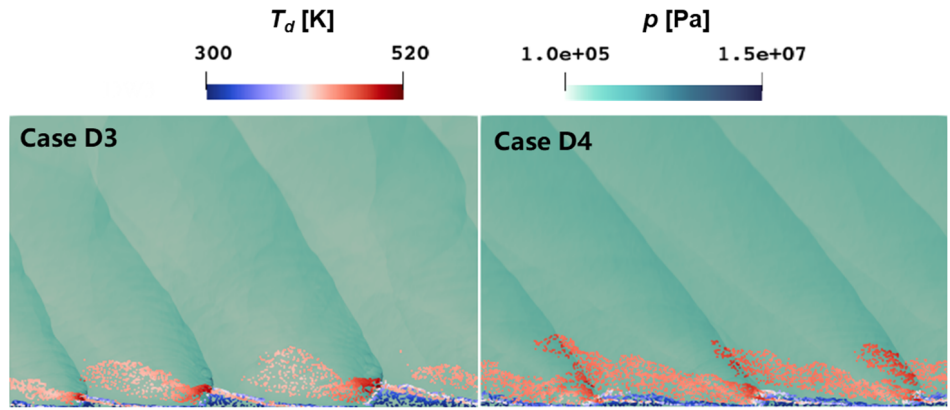

Figure 19 Stabilized two-phase RDW flow fields under the three-wave mode. Contours of pressure and droplet temperature.

It should be noted that despite both Cases D3 and D4 propagating with a three-wave structure, differences in the distribution of fuel droplets persist. In both cases, the droplets in the full-refill region fail to completely consume after being swept by the detonation wave and continue moving towards the post-RDW region. However, Fig. 19 indicates that in Case D3, the travel of droplets in the post-detonation region terminates before reaching the OSW of the subsequent detonation wave. On the other hand, non-evaporated droplets in Case D4 reach beyond the OSW of the succeeding detonation wave. To delve deeper into the interaction between droplets and the OSW, we examine the flow field of Case D4 from various perspectives (Fig. 20). The droplets experience further heating to high temperatures upon traversing the OSW. However, the distribution of ethanol vapor near the OSW suggests that there is no significant change in the volume of ethanol vapor before and after passing through the OSW, indicating that the droplets undergo negligible evaporation despite being heated by the OSW. This observation is further supported by the plots of the heat release rate and the distribution of oxygen.

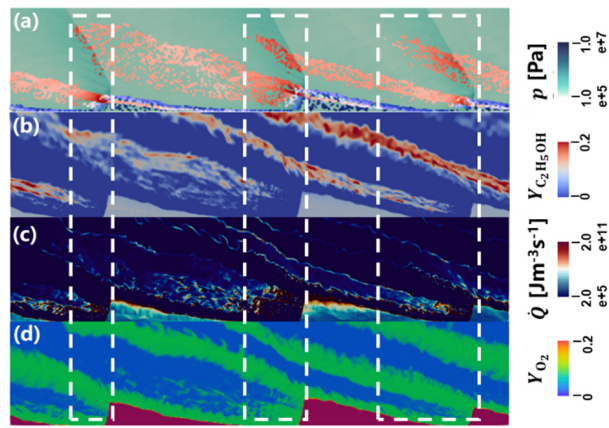

Figure 20 Contours of (a) pressure, (b) ethanol vapor mass fraction, (c) heat release rate, and (d) oxygen mass fraction of the flow field of Case D4.

## 4. Concluding remarks

As an attempt to investigate the viability of utilizing biofuel as an alternative to conventional fossil fuels in the RDE, this study provides some insights into the behavior of the two-phase RDW related to the evaporation process. To establish self-sustained ethanol-fueled RDWs, liquid ethanol is partially pre-vaporized, and droplets with diameters ranging from 5 to 25 μm are injected. The air is pre-heated at temperatures ranging from 900 K to 1200 K to facilitate the evaporation of the liquid fuel.

Under the conditions of an inlet total temperature of 900 K and a droplet diameter of 5 μm, the primary structure of the two-phase ethanol-air RDW closely resembles that of the premixed ethanol vapor-air RDW, as the droplets are instantly evaporated and consumed by the detonation front. When the diameter of injected ethanol droplets exceeds 10 μm, the interactions between the droplets and RDW become a two-stage process: after the droplets are swept by the detonation front, secondary evaporation and reactions continue right after the detonation front. This is attributed to the prolonged duration required for the completion of droplet evaporation, leading to subsequent secondary-stage evaporation and reactions. For $d_0 > 15$ μm, the droplets can travel a long distance in the post-detonation zone, and the completion of reactions no longer concentrates near the detonation front.

Furthermore, this study investigates the effect of inlet total temperature on the evaporation and pre-ignition of liquid droplets. Droplets exhibit varying evaporation rates at elevated temperatures ranging from 1000 K to 1200 K. In the refill zone, a certain fraction of ethanol vapor undergoes pre-ignition before being swept by the RDW. The presence of local explosions and shock waves leads to collisions and the generation of new detonation waves. Stable double- or three-wave flow fields are formed when the droplet diameter increases to 10-25 μm at an inlet total temperature of 1200 K, whereas the flow field fails to achieve re-stabilization with a droplet diameter of 5 μm due to intensive pre-ignition and persistent generation, extinction, and collisions of detonation waves.

**REFERENCES**

[1] V. Raman, S. Prakash, M. Gamba, Nonidealities in Rotating Detonation Engines, Annual Review of Fluid Mechanics, 55 (2023) 639-674.

[2] P. Wolański, Detonative propulsion, Proceedings of the Combustion Institute, 34 (2013) 125-158.

[3] J.Z. Ma, M.Y. Luan, Z.J. Xia, J.P. Wang, S.J. Zhang, S.B. Yao, B. Wang, Recent progress, development trends, and consideration of continuous detonation engines, AIAA Journal, 58 (2020) 4976-5035.

[4] T.W. Teasley, T.M. Fedotowsky, P.R. Gradl, B.L. Austin, S.D. Heister, Current State of NASA Continuously Rotating Detonation Cycle Engine Development, in: AIAA SCITECH 2023 Forum, 2023.

[5] K. Goto, K. Matsuoka, K. Matsuyama, A. Kawasaki, H. Watanabe, N. Itouyama, K. Ishihara, V. Buyakofu, T. Noda, J. Kasahara, A. Matsuo, I. Funaki, D. Nakata, M. Uchiumi, H. Habu, S. Takeuchi, S. Arakawa, J. Masuda, K. Maehara, T. Nakao, K. Yamada, Space Flight Demonstration of Rotating Detonation Engine Using Sounding Rocket S-520-31, Journal of Spacecraft and Rockets, 60 (2023) 273-285.

[6] F.A. Bykovskii, S.A. Zhdan, E.F. Vedernikov, Continuous spin detonations, Journal of Propulsion and Power, 22 (2006) 1204-1216.

[7] M.L. Fotia, J. Hoke, F. Schauer, Experimental study of performance scaling in rotating detonation engines operated on hydrogen and gaseous hydrocarbon fuel, 20th AIAA International Space Planes and Hypersonic Systems and Technologies Conference, 2015, (2015) 1-22.

[8] Z. Ma, S. Zhang, M. Luan, S. Yao, Z. Xia, J. Wang, Experimental research on ignition, quenching, reinitiation and the stabilization process in rotating detonation engine, International Journal of Hydrogen Energy, 43 (2018) 18521-18529.

[9] V.F. Nikitin, E.V. Mikhalchenko, Safety of a rotating detonation engine fed by acetylene - oxygen mixture launching stage, Acta Astronautica, 194 (2022) 496-503.

[10] D. Schwer, K. Kailasanath, Fluid dynamics of rotating detonation engines with hydrogen and hydrocarbon fuels, Proceedings of the Combustion Institute, 34 (2013) 1991-1998.

[11] J. Kindracki, P. Wolański, Z. Gut, Experimental research on the rotating detonation in gaseous fuels–oxygen mixtures, Shock Waves, 21 (2011) 75-84.

[12] N.N. Smirnov, V.F. Nikitin, L.I. Stamov, E.V. Mikhalchenko, V.V. Tyurenkova, Rotating detonation in a ramjet engine three-dimensional modeling, Aerospace Science and Technology, 81 (2018) 213-224.

[13] J. Zhou, F. Song, Y. Wu, S. Xu, X. Yang, P. Cheng, Y. Li, Investigation of pressure gain characteristics for kerosene-hot air RDE, Combustion and Flame, 247 (2023) 126102.

[14] M.H. Zhao, K. Wang, Y.Y. Zhu, Z.C. Wang, Y. Yan, Y.J. Wang, W. Fan, Effects of the exit convergent ratio on the propagation behavior of rotating detonations utilizing liquid kerosene, Acta Astronautica, 193 (2022) 35-43.

[15] Q. Zheng, H. Meng, C. Weng, Y. Wu, W. Feng, M. Wu, Experimental research on the instability propagation characteristics of liquid kerosene rotating detonation wave, Defence Technology, 16 (2020) 1106-1115.

[16] J. Han, Q. Bai, S. Zhang, C. Weng, Experimental study on propagation mode of rotating detonation wave with cracked kerosene gas and ambient temperature air, Physics of Fluids, 34 (2022) 075127.

[17] S. Xue, Z. Ying, H. Ma, C. Zhou, Experimental Investigation on Two-Phase Rotating Detonation Fueled by Kerosene in a Hollow Directed Combustor, Frontiers in Energy Research, 10 (2022).

[18] G. Xu, Y.W. Wu, Q. Xiao, C.W. Ding, Y.Q. Xia, Q. Li, C.S. Weng, Characterization of wave modes in a kerosene-fueled rotating detonation combustor with varied injection area ratios, Applied Thermal Engineering, 212 (2022) 118607.

[19] J. Kindracki, Experimental research on rotating detonation in liquid fuel-gaseous air mixtures, Aerospace Science and Technology, 43 (2015) 445-453.

[20] F.A. Bykovskii, S.A. Zhdan, E.F. Vedernikov, Continuous Detonation of the Liquid Kerosene-Air Mixture with Addition of Hydrogen or Syngas, Combustion, Explosion and Shock Waves, 55 (2019) 589-598.

[21] S.Y. Huang, J. Zhou, S.J. Liu, H.Y. Peng, X.Q. Yuan, Continuous rotating detonation engine fueled by ammonia, Energy, 252 (2022).

[22] Z. Sun, Y. Huang, Z. Luan, S. Gao, Y. You, Three-dimensional simulation of a rotating detonation engine in ammonia/hydrogen mixtures and oxygen-enriched air, International Journal of Hydrogen Energy, 48 (2022) 4891-4905.

[23] F. Wang, Q. Liu, C. Weng, On the feasibility and performance of the ammonia/hydrogen/air rotating detonation engines, Physics of Fluids, 35 (2023).

[24] K. Yoneyama, K. Ishihara, S. Ito, H. Watanabe, N. Itouyama, A. Kawasaki, K. Matsuoka, J. Kasahara, A. Matsuo, I. Funaki, Experimental Clarification on Detonation Phenomena of Liquid Ethanol Rotating Detonation Combustor, in: AIAA SCITECH 2022 Forum, 2022.

[25] T. Sato, K. Ishihara, K. Yoneyama, S. Ito, N. Itouyama, H. Watanabe, A. Kawasaki, K. Matsuoka, J. Kasahara, A. Matsuo, I. Funaki, Experimental Research on Thrust Performance of Rotating Detonation Engine with Liquid Ethanol and Gaseous Oxygen, in: AIAA AVIATION 2022 Forum, 2022.

[26] S. Yao, X. Tang, W. Zhang, Structure of a heterogeneous two-phase rotating detonation wave with ethanol–hydrogen–air mixture, Physics of Fluids, 35 (2023) 031712.

[27] S. Yao, C. Guo, W. Zhang, Effects of droplet evaporation on the flow field of hydrogen-enhanced rotating detonation engines with liquid kerosene, International Journal of Hydrogen Energy, (2023).

[28] Q. Meng, M. Zhao, H. Zheng, H. Zhang, Eulerian-Lagrangian modelling of rotating detonative combustion in partially pre-vaporized n-heptane sprays with hydrogen addition, Fuel, 290 (2021).

[29] W. Sutherland, LII. The viscosity of gases and molecular force, The London, Edinburgh, and Dublin Philosophical Magazine and Journal of Science, 36 (1893) 507-531.

[30] R.A. Svehla, Estimated viscosities and thermal conductivities of gases at high temperatures, in, NASA, 1962.

[31] M. Zhao, M.J. Cleary, H. Zhang, Combustion mode and wave multiplicity in rotating detonative combustion with separate reactant injection, Combustion and Flame, 225 (2021) 291-304.

[32] Z. Huang, M. Zhao, Y. Xu, G. Li, H. Zhang, Eulerian-Lagrangian modelling of detonative combustion in two-phase gas-droplet mixtures with OpenFOAM: Validations and verifications, Fuel, 286 (2021) 119402.

[33] B. Abramzon, W.A. Sirignano, Droplet vaporization model for spray combustion calculations, International Journal of Heat and Mass Transfer, 32 (1989) 1605-1618.

[34] V.B. Betelin, N.N. Smirnov, V.F. Nikitin, V.R. Dushin, A.G. Kushnirenko, V.A. Nerchenko, Evaporation and ignition of droplets in combustion chambers modeling and simulation, Acta Astronautica, 70 (2012) 23-35.

[35] N.N. Smirnov, V.B. Betelin, A.G. Kushnirenko, V.F. Nikitin, V.R. Dushin, V.A. Nerchenko, Ignition of fuel sprays by shock wave mathematical modeling and numerical simulation, Acta Astronautica, 87 (2013) 14-29.

[36] V.V. Tyurenkova, Non-equilibrium diffusion combustion of a fuel droplet, Acta Astronautica, 75 (2012) 78-84.

[37] R.S. Miller, K. Harstad, J. Bellan, Evaluation of equilibrium and non-equilibrium evaporation models for many-droplet gas-liquid flow simulations, International Journal of Multiphase Flow, 24 (1998) 1025-1055.

[38] H. Wen, W. Fan, S. Xu, B. Wang, Numerical study on droplet evaporation and propagation stability in normal-temperature two-phase rotating detonation system, Aerospace Science and Technology, 138 (2023).

[39] S. Jin, H. Zhang, N. Zhao, H. Zheng, Simulations of rotating detonation combustion with in-situ evaporating bi-disperse n-heptane sprays, Fuel, 314 (2022).

[40] M. Zhao, H. Zhang, Rotating detonative combustion in partially pre-vaporized dilute n-heptane sprays: Droplet size and equivalence ratio effects, Fuel, 304 (2021).

[41] W.E. Ranz, W.R. Marshall, Evaporation from drops: Part 1, Chemical Engineering Progress, 48 (1952) 141-146.

[42] W.E. Ranz, W.R. Marshall, Evaporation from drops: Part 2, Chemical Engineering Progress, 48 (1952) 173-180.

[43] K.K. Kuo, R. Acharya, Fundamentals of turbulent and multiphase combustion, John Wiley & Sons, 2012.

[44] C.K. Westbrook, F.L. Dryer, Simplified Reaction Mechanisms for the Oxidation of Hydrocarbon Fuels in Flames, Combustion Science and Technology, 27 (1981) 31-43.

[45] H.G. Weller, G. Tabor, H. Jasak, C. Fureby, A tensorial approach to computational continuum mechanics using object-oriented techniques, Computers in Physics, 12 (1998).

[46] A. Kurganov, E. Tadmor, New High-Resolution Central Schemes for Nonlinear Conservation Laws and Convection–Diffusion Equations, Journal of Computational Physics, 160 (2000) 241-282.

[47] K. Wu, S. Zhang, M. Luan, J. Wang, Effects of flow-field structures on the stability of rotating detonation ramjet engine, Acta Astronautica, 168 (2020) 174-181.

[48] N.N. Smirnov, V.F. Nikitin, L.I. Stamov, E.V. Mikhalchenko, V.V. Tyurenkova, Three-dimensional modeling of rotating detonation in a ramjet engine, Acta Astronautica, 163 (2019) 168-176.

[49] X.M. Tang, J.P. Wang, Y.T. Shao, Three-dimensional numerical investigations of the rotating detonation engine with a hollow combustor, Combustion and Flame, 162 (2015) 997-1008.

[50] N. Jourdaine, N. Tsuboi, K. Ozawa, T. Kojima, A.K. Hayashi, Three-dimensional numerical thrust performance analysis of hydrogen fuel mixture rotating detonation engine with aerospike nozzle, Proceedings of the Combustion Institute, 37 (2019) 3443-3451.

[51] V.B. Betelin, A.G. Kushnirenko, N.N. Smirnov, V.F. Nikitin, V.V. Tyurenkova, L.I. Stamov, Numerical investigations of hybrid rocket engines, Acta Astronautica, 144 (2018) 363-370.

[52] B. Muralidharan, S. Menon, Simulation of moving boundaries interacting with compressible reacting flows using a second-order adaptive Cartesian cut-cell method, Journal of Computational Physics, 357 (2018) 230-262.

[53] K. Kailasanath, Liquid-fueled detonations in tubes, Journal of Propulsion and Power, 22 (2006) 1261-1268.

[54] X.-Y. Liu, M.-Y. Luan, Y.-L. Chen, J.-P. Wang, Propagation behavior of rotating detonation waves with premixed kerosene/air mixtures, Fuel, 294 (2021) 120253.

[55] P. Diakow, M. Cross, G. Ciccarelli, Detonation characteristics of dimethyl ether and ethanol–air mixtures, Shock Waves, 25 (2015) 231-238.

[56] N.N. Smirnov, V.B. Betelin, V.F. Nikitin, L.I. Stamov, D.I. Altoukhov, Accumulation of errors in numerical simulations of chemically reacting gas dynamics, Acta Astronautica, 117 (2015) 338-355.

[57] M. Zhao, H. Zhang, Origin and chaotic propagation of multiple rotating detonation waves in hydrogen/air mixtures, Fuel, 275 (2020) 117986.